\documentclass[amssymb,epsfig,color,11pt]{article}
\usepackage{epsfig}
\usepackage{amsfonts}
\usepackage{amssymb}
\usepackage{indentfirst}
\usepackage{color}
\usepackage{mathrsfs}
\def\bc{\begin{center}}

\def\ec{\end{center}}
\def\be{\begin{eqnarray}}
\def\ee{\end{eqnarray}}

\definecolor{dyellow}{rgb}{1.,0.8,.0}
\definecolor{myblue}{rgb}{.1,.1,.7}
\definecolor{dcyan}{rgb}{.0,.6,.6}
\definecolor{dmagenta}{rgb}{0.6,0.0,0.6}
\definecolor{brown}{rgb}{0.6,0.2,0.}
\definecolor{darkblue}{rgb}{.0,.0,0.5}
\definecolor{darkred}{rgb}{0.75,0.0,0.0}
\definecolor{orange}{rgb}{1.,.6,.0}
\definecolor{dorange}{rgb}{0.8,.4,.0}
\definecolor{darkgreen}{rgb}{0.0,0.6,0.0}
\definecolor{purple}{rgb}{.4,.0,.4}
\begin{document}
\baselineskip=16pt
\newcommand{\omits}[1]{}

%\hfill{\bf USTC-ICTS-07-05} \vspace{1.0cm}

\begin{center}
{\Large \bf Extremely high energy cosmic ray as the probe of new physics}\\
\vspace{1cm}
Shao-Xia Chen\footnote{ruxanna@sdu.edu.cn}\\
{\em School of Space Science and Physics,}\\
 {\em Shandong University at Weihai,}\\
{\em Weihai, Shandong 264209, China}\\
Rong Hu\\
{\em School of Mechanical Engineering}\\
{\em Beijing Technology and Business University}\\
{\em Beijing 100048, China}\\
\end{center}

\begin{abstract}
We explore the influence of new physics on extremely high energy
cosmic ray (EXCER) particles. In particular, we devote our mind to
one example of new physics, unparticle stuff, on one specific
process that EXECRs participate in, photopion production of the
EXECR nucleon with 2.7K cosmic microwave background radiation
(CMBR). Through computing the differential cross section of virtual
exchange of unparticle in the $p+\gamma\to p+ \pi^0$ process, we
acquire the general consequence of new physics on the EXECR
propagation. It is astonishing but reasonable that due to the
lowness of interaction energy $\sqrt{s}$, the new physics will play
a nearly negligible role in EXECR interaction.
\end{abstract}

\vspace{1.0cm}


\vspace{1.0cm}

Since cosmic ray was discovered in 1912 by Hess \cite{hess}, it
became one powerful tool in particle physics in that it had been an
important method to discover new particles such as positron, muon
and $\pi$ before accelerators were constructed in 1950s. Nowadays
owing to its stupendous energy which is greater by about seven
orders of magnitude than that the particles in the terrestrial
laboratories can obtain, EXECR that generally is termed as the
cosmic ray with energy in excess of $10^{19}~$eV, is always deemed
as the most hopeful probe to explore various candidates of so-called
new physics. In the past decades there are really tremendous efforts
\cite{VLI} on testing new physics with EXECR, especially testing
manifestation of VLI induced from various quantum gravity models
using the GZK feature of EXECR.

It had been believed that we could obtain the cosmic ray particles
with no upper limit of energy and the reason that they hadn't been
detected so far was that their flux was too low to our detectors.
However, shortly after the discovery of CMBR in 1966, Greisen
\cite{G}, Zatsepin and Kuzmin \cite{ZK} pointed out that the cosmic
ray nucleon will interact with the background photon $N+\gamma\to
N+\pi$, and lose about 50\% energy, respectively. This process will
lead to the GZK feature that the spectrum of cosmic ray will steepen
at the GZK predicted energy, about $5\times10^{19}$eV, for the giant
energy loss of cosmic ray particle before it arrives at the
atmosphere.

There are three observatories on EXECR, i.e., the Akeno Giant Air
Shower Array (AGASA), the High Resolution Fly's Eye Cosmic Ray
Detector (HiRes) and the southern Pierre Auger Observatory (PAO).
Nevertheless, there are somewhat controversial on the observations
of the EXECR particles among them. The results of AGASA \cite{AGASA}
indicate an obvious absence of GZK feature, while HiRes
\cite{Hires-GZK} and the southern PAO \cite{PAO-GZK} do observe the
GZK cutoff. What's more, HiRes \cite{Hires-com} shows the
composition of EXECR transfers from heavy nuclei to protons at about
$3\times10^{15}$ eV, and PAO \cite{PAO-com} favors the results that
the composition above $10^{19}$eV is mostly heavy nuclei. In
addition, a correlation of the EXECR events with nearby Active
Galactic Nuclei (AGN) was reported \cite{AGN-cor} according to the
early data of PAO, however, HiRes\cite{hires-cor} and the recent PAO
data \cite{PAO-cor} reveal no significant correlation of EXECR with
any celestial objects.

In general, new physics are classified into two categories. One is
motivated by quantizing gravity or unifying the four kinds of
fundamental interactions, such as extra dimension, loop quantum
gravity, and the recently proposed entropic force \cite{verlinde}.
The other is particle physics models extending on the basis of the
Standard Model (SM) SU(3)$\bigotimes$SU(2)$\bigotimes$U(1) of
particle physics, for instance, supersymmetry theory, little Higgs
models, and recent unparticle physics \cite{Georgi1}. The previous
pursuits are mainly on applying a very small violation of Lorentz
invariance (VLI) related to the former kind of new physics to
illustrate the absence of GZK cutoff that AGASA observed. The
situation changed crucially after the HiRes and PAO announced their
results and confirmed the predicted GZK cutoff, which seemingly does
not entail VLI. However, Stecker \cite{stecker-new} recently
reexcited the interest in VLI on EXECR through proposing another
scenario in which VLI will coexist with the HiRes and PAO
experimental data.

Nowadays there are a large number of new physics models and theories
and a lot of candidates of EXECR particle such as nucleon and heavy
nuclei. The extension of SM (that is, the latter kind of new
physics) usually predicts new ``particles" and couplings that may
contribute to the interactions that EXECR particles participate in
during their propagation and generate the observable signature in
the EXECR spectrum. In this work, we examine the possible effect of
the extension of SM on these interactions. For the sake of
simplicity and typicality, we take the recent unparticle physics and
nucleon as the examples of the new physics and EXECR particle,
respectively, and derive some common results for the impact of new
physics on the EXECR propagation in the end.

In 2007, Georgi \cite{Georgi1} proposed the existence of unparticle
which is a scale invariant sector with a non-trivial infrared
fixed-point. In the following years, a number of papers appear and
cover a lot applications of unparticle physics, such as collider
physics \cite{collider}, $CP$ violation \cite{CP}, Higgs physics
\cite{higgs}, cosmology \cite{cos}, and supersymmetry \cite{susy},
etc., and also focus on many fundamental issues in unparticle
physics \cite{funda}.

It is evident that the unparticle can only play a role in the
neutral pion $\pi^0$ photoproduction $p+\gamma\to p+\pi^0$ but not
in the similar charged pion $\pi^\pm$ photoproduction $p+\gamma\to
n+\pi^+$ and $n+\gamma\to p+\pi^-$ on the premise that the
unparticle stuff does not possess the quantum number of gauge group,
such as the electric charge $Q$. The process can only occur through
the $t$-channel for axial vector unparticle operator $O_{\cal
U}^\mu$, and the effective couplings among $\gamma$, $\pi^0$, and
vector unparticle operator $O_{\cal U}^\mu$ is
\begin{equation}\label{L}
  L_{\gamma\pi{\cal U}}=\displaystyle\frac{ie\lambda_1}{4\Lambda_{\cal U}^{d_{\cal
  U}}}\epsilon_{\mu\nu\rho\sigma}F^{\mu\nu}O_{\cal
  U}^{\rho\sigma}\Pi~,
\end{equation}
where $\Pi$ is the pion field, $\Lambda_{\cal U}$ is the energy
scale at which scale invariance emerges,
$\epsilon_{\mu\nu\rho\sigma}$ is the totally antisymmetric tensor,
$F^{\mu\nu}=\partial^\mu A^\nu-\partial^\nu A^\mu$ is the
electromagnetic field strength, and $O_{\cal
U}^{\mu\nu}:=\partial^\mu O_{\cal U}^\nu-\partial^\nu
  O_{\cal U}^\mu$. Combined (\ref{L}) and the effective couplings among two protons and
vector unparticle operator $O_{\cal U}^\mu$ \cite{collider}, we can
acquire the spin-averaged amplitudes squared is deduced as
\begin{equation}\label{amplitudes}
\begin{array}{rcl}
\overline{ |{\cal M}|^2}&=&
\displaystyle\frac{8m^6-2(4s+t+4u)m^4+2(s+u)(2m^2+m_\pi^2)m^2-t(s^2+u^2)}{|t|^{3}}\left(\displaystyle\frac{|t|}
{\Lambda_{\cal
U}^{2}}\right)^{2d_{\cal U}-1}\\
[0.8cm]&&\times\displaystyle\frac{ e^2\lambda_1^2\lambda_2^2
Z^2_{d_{\cal U}}} {4}~,~~~~~~ Z_{d_{\cal
U}}=\displaystyle\frac{8\pi^{5/2}}{\sin(d_{\cal
U}\pi)(2\pi)^{2d_{\cal U}}} \displaystyle\frac{\Gamma(d_{\cal
U}+1/2)}{\Gamma(d_{\cal U}-1)\Gamma(2d_{\cal U})}~,
    \end{array}
\end{equation}
where $m$ is the mass of proton, $m_\pi$ is the mass of $\pi^0$,
$\lambda_2$ is the coupling constant among two protons and
unparticle operator $O_{\cal U}^\mu$, and $d_{\cal U}$ is the scale
dimension of the unparticle operator $O_{\cal U}$. The differential
cross section in the center of mass (cm) frame is derived as follows
\begin{equation}\label{sigma}
\displaystyle\frac{d\sigma}{d\Omega}= \displaystyle\frac{p_1
 }{64\pi^2 Ek(E+k)}\overline{|{\cal M}|^2}~,
\end{equation}
where $k$ is the energy of incident photon, $E$ is the energy of the
incident proton, and $p_1$ is the magnitude of momentum vector of
the outgoing proton.

Now we will come to the details but key points. The energy spectrum
of cosmic ray particles can extend to the highest $10^{21}$eV and
the typical energy $\epsilon$ of CMBR photons is about $10^{-3}$eV.
Thus in the earth frame the 4-momentum of EXECR proton can be taken
as $p_{\rm p}=(E_1,0,0,p_1)$ and that of CMBR photon will be
$p_\gamma=(E_\gamma,0,0,-E_\gamma)$. The Mandelstam variable $s$
that is the square of cm energy of the interaction between an EXECR
proton and a CMBR photon is $s=(p_{\rm p}+p_\gamma)^2=m_{\rm
p}^2+2(E_1+p_1)E_\gamma\sim$GeV$^2$ owing to the tininess of the
typical energy of CMBR photon. {\it The total center-of-mass energy
is only in the GeV order, which can be achieved in the terrestrial
laboratories and new physics will not appear at this energy scale
much below TeV!}

Let's continue to calculate the cross section and compare it with
that derived in the Standard Model without new physics to present
some quantitative and straightforward results. The unknown
quantities on the right-hand of Eq.(\ref{sigma}) can be derived from
the earth frame to the cm frame via the Lorentz invariance of
Mandelstam variable $s$, $t$ and $u$. What's more, the energy scale
of the interaction can be determined by the energy $k$ of incident
photon in the cm frame due to the following fact: the 4-momentum of
the incident proton and photon are separately $p^{\rm cm}_{\rm
p}=(E,0,0,k)$ and $p^{\rm cm}_\gamma=(k,0,0,-k)$, which leads to the
energy scale of the interaction is
\begin{equation}\label{s-cm}
 \sqrt{s}=\sqrt{(p^{\rm cm}_{\rm p}+p^{\rm cm}_\gamma)^2}=E+k=\sqrt{m_{\rm
 p}^2+k^2}+k~.
 \end{equation}
As a consequence, $k$ will also at the GeV scale in the case
$\sqrt{s}\sim$ GeV.

In order to show the influence of new physics on the differential
cross section clearly, it is necessary to present some curves of
$d\sigma/d\Omega$ versus the pion cm angle with different $d_{\cal
U}$ for several $k$ in the case $\Lambda_{\cal U}$ and
$\lambda_1=\lambda_2$ are in the reasonable range. For the
convenience to compare with the experimental data \cite{cross-data},
the energy of incident photon in the cm frame are taken as $k=0.26$
GeV (corresponding to $\sqrt{s}=1.233$ GeV) and $k=0.32$ GeV
(corresponding to $\sqrt{s}=1.311$ GeV) in Fig.1 and Fig.2,
respectively, which are same as those in \cite{cross-data}. The
other parameters are taken as $\Lambda_{\cal U}=1~$TeV and
$\lambda_1=\lambda_2=1.0$.
\begin{center}
\includegraphics[totalheight=9.5cm]{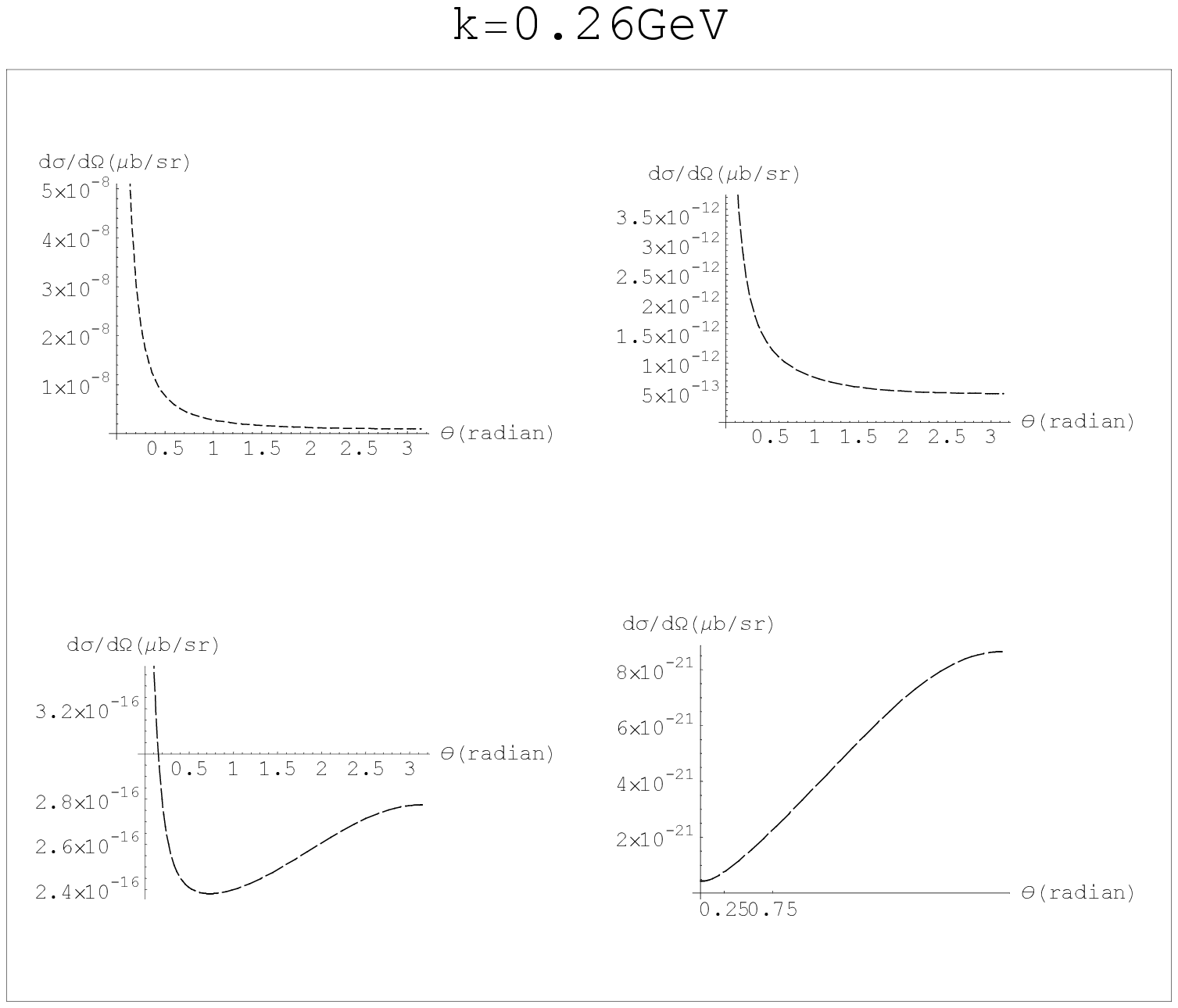}
\end{center}
{\bf Fig.1} The differential cross section $d\sigma/d\Omega$ in
units of $\mu$b vs the pion cm angle $\theta$ in the case $k=0.26$
GeV for $d_{\cal U}=1.1,~1.3,~1.5,~1.8$. The dashes get longer as
$d_{\cal U}$ increases.
\begin{center}
\includegraphics[totalheight=9.5cm]{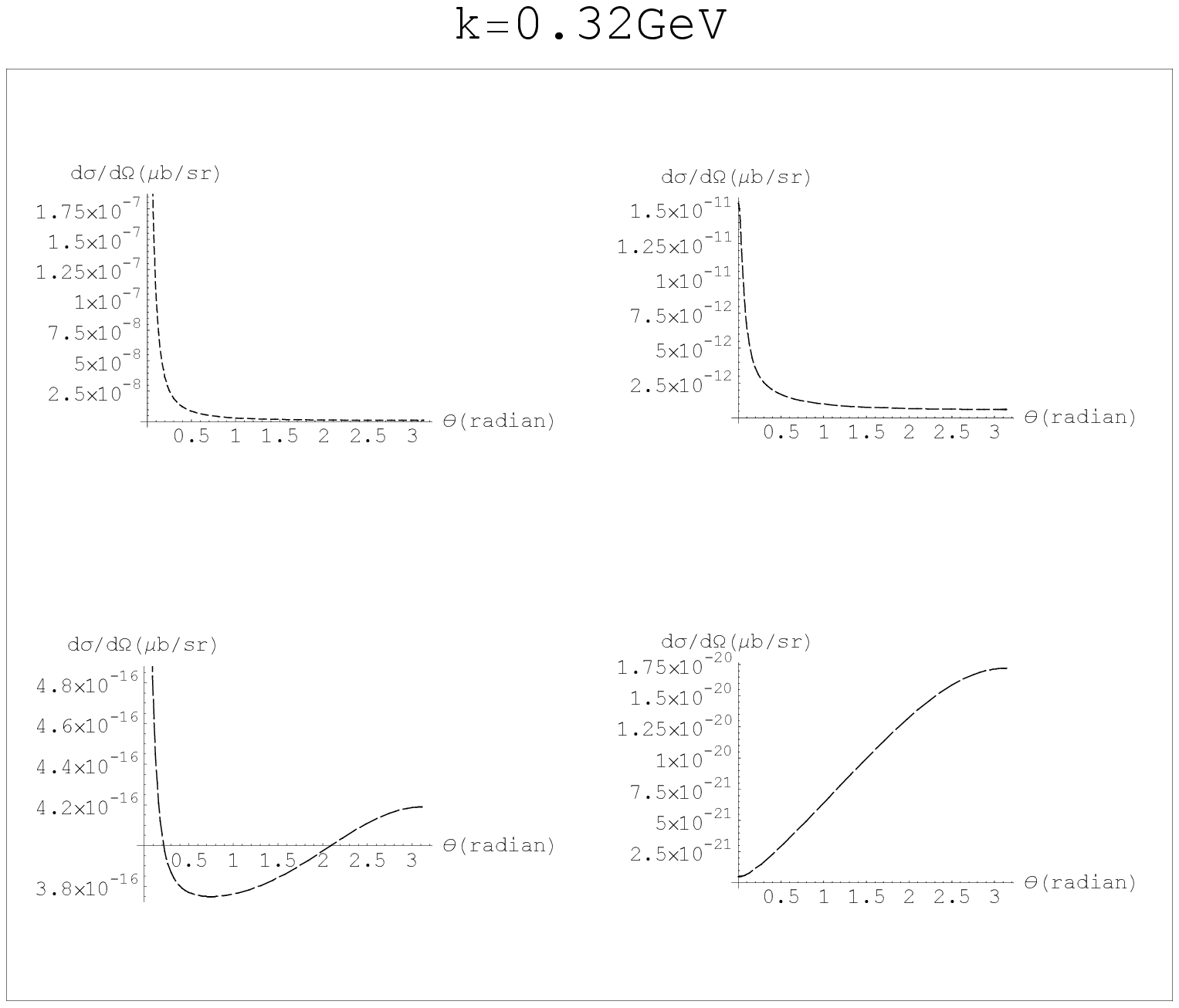}\\
{\bf Fig.2} Same as Fig.1 except $k=0.32$ GeV.
\end{center}
In terms of Fig.1 and Fig.2, it is astonishing but reasonable that
the contribution from the unparticle stuff can really be overlooked
compared to the data of differential cross section of pion
photoproduction with magnitude of order of tens of $\mu$b
\cite{cross-data} provided that $\Lambda_{\cal U}$ is in the proper
range, no matter the way $d_{\cal U}$ changes. The key point to
comprehend this conclusion is as follows. Although EXECR is seemed
as the most hopeful window to probe the new physics, but, it is not
stressed so far that the interaction energy $\sqrt{s}$ of the pion
meson photoproduction is about 1 GeV at which scale nearly all new
physics do not manifest themselves.

In addition, the lowness of $\sqrt{s}$ is not only limited to the
pion photoproduction but a general feature of all processes that
propagating EXECR particles participate in (such as the
photodisintegration of EXECR nuclei with CMBR and IR background
photons \cite{nuclei}, the pair-production of UHE photons off the
background photons: $\gamma\gamma_{\rm B}\to e^+e^-$, and inverse
Compton scattering (ICS) of the electrons (positrons) on the
background photons, etc.) owing to the lowness of the typical
energies of various background radiations. Consequently, it is
general and justified that new physics will have minute effect on
the propagating EXECRs' interactions except the possible alteration
of their spectra from VLI motivated from the new physical model or
theory based on the previous illustration and combined with our
previous results \cite{our}.

In the final let us close with some comments
\begin{enumerate}
  \item For the non-conventional candidates of EXECR particle, such
  as neutrino and exotic particles, similar effects can also be
  discussed. However, the PAO presents the upper limit on detected
  photon flux \cite{PAO-photon}, which disfavors most exotic candidates, and the
  observed profile of extensive air shower is somewhat
  contradictory with that neutrino produces. Hence our subject just
  involves the conventional EXECR particles namely nucleon, nuclei and
  $e^\pm$, $\gamma$ \cite{our}.
\item It is obvious that our investigation is different from the
work motivated from the VLI and our results are distinct reasonably.
Moreover, there is another important distinction between our and
others' researches. There are no special frame in our derivation due
to the Lorentz invariance of the Mandelstam variables hence our
analysis are frame-independent. In contrast, the extent of VLI is
frame-dependent and the conclusions from VLI are also
frame-dependent which are often drawn in the Earth frame.
\end{enumerate}
In conclusion, the two categories of new physics have distinct
significance on the interactions of EXECR with various background
radiations: Some of the former that predict the VLI will affect the
observed spectrum of EXECR to some extent \cite{stecker-new}, and
the latter almost only play a negligible role in the same
interactions.

\section*{Acknowledgments}
The work is supported in part by the Science Foundation of Shandong
University at Weihai under Grant No. 0000507300020.

%%%%%%%%%%%%%%%%%%%%%%%%%%%%%%%%%%%%%%%%%%%%%%%%%%%%%%%%%%%%%%%%%%%%%%%%%%

%\end{CJK*}

\begin{thebibliography}{999}
%%%%%%%%%%%%%%%%%%%%%%%%%%Hess
\bibitem{hess} V. F. Hess, Phys. Z. {\bf 13}, (1912) 1084.

%%%%%%%%%%%%%%%%%%%%%new physics and EXECR
\bibitem{VLI} S. Coleman and S. L. Glashow, Phys. Lett. B {\bf 405},
(1997) 249; S. Coleman and S. L. Glashow, Phys. Rev. D {\bf 59},
(1999) 116008; F. W. Stecker and S. Glashow, Astropart. Phys.
 {\bf 16}, (2001) 97; N. Gupta, Phys. Lett. B {\bf 580},
(2004) 103; G. Amelino-Camelia, Nature {\bf 408}, (2000) 661.

%%%%%%%%%%%%%%%%%%GZK
\bibitem{G} K. Greisen, Phys. Rev. Lett. {\bf 16}, (1966) 748.

\bibitem{ZK} G. T. Zatsepin and V. A. Kuzmin, JETP Lett. {\bf 4},
(1966) 78.

%%%%%%%%%%%%%%%%%%%%%%GZK observation
\bibitem{AGASA} M. Takeda {\it et al}., Astropart. Phys. {\bf 19}, (2003)
447.

\bibitem{Hires-GZK} R. U. Abbasi {\it et al}., Phys. Rev. Lett. {\bf 100}, (2008)
101101.

\bibitem{PAO-GZK} J. Abraham {\it et al}., Phys. Rev. Lett. {\bf 101}, (2008) 061101;
{\it ibid}, Phys. Lett. B {\bf 685}, (2010) 239.

%%%%%%%%%%%%%%%%%%%%%%%%%composition
\bibitem{Hires-com} R. U. Abbasi {\it et al}., Phys. Rev. Lett. {\bf 104}, (2010)
091101.

\bibitem{PAO-com} J. Abraham {\it et al}., Phys. Rev. Lett. {\bf 104}, (2010)
091101.
%%%%%%%%%%%%%%%%%%%%%%%%EXECR-direction

\bibitem{AGN-cor} J. Abraham {\it et al}., Science {\bf 318}, (2007) 938.

\bibitem{hires-cor} R. U. Abbasi {\it et al}., Astropart. Phys. {\bf 30}, (2008)
175.

\bibitem{PAO-cor} J. Abraham {\it et al}., Proc. 31st Int. Cosmic Ray Conf., (2009).
%%%%%%%%%%%%%%%%%%%%%%entropic force
\bibitem{verlinde} E. P. Verlinde, arXiv: 1001.0785.

%%%%%%%%%%%%%%%%%%%%%%%%%%%unparticle

\bibitem{Georgi1} H. Georgi, Phys. Rev. Lett. {\bf 98}, (2007) 221601.

%%%%%%%%%%%%%%%%%%%%%%%%%%%%%Stecker
\bibitem{stecker-new} F. W. Stecker and S. T. Scully, New J. Phys.
{\bf 11}, (2009) 085003.

%\bibitem{BZ} T. Banks and A. Zaks, Nucl. Phys. B {\bf 196}, (1982) 189.

%\bibitem{Georgi2} H. Georgi, Phys. Lett. B {\bf 650}, (2007) 275.

\bibitem{collider} K. Cheung, W. Y. Keung and T. C. Yuan, Phys. Rev. Lett. {\bf 99}, (2007)
051803; {\it ibid}, Phys. Rev. D {\bf 76}, (2007) 055003.

%%%%%%%%%%%%%%%%%%CP

\bibitem{CP} C. H. Chen and C. Q. Geng, Phys. Rev. D {\bf 76}, (2007)
115003; C. H. Chen and C. Q. Geng, Phys. Rev. D {\bf 76}, (2007)
036007; R. Zwicky, Phys. Rev. D {\bf 77}, (2008) 036004; C. S. Huang
and X. H. Wu, Phys. Rev. D {\bf 77}, (2008) 075014; R. Mohanta and
A. K. Giri, Phys. Rev. D {\bf 76}, (2007) 057701.

%%%%%%%%%%%%%%%%%%

%%%%%%%%%%%%%%%%%%Higgs

\bibitem{higgs} M. X. Luo and G. H. Zhu, Phys. Lett. B {\bf 659}, (2008) 341;
P. J. Fox, A. Rajaraman, and Y. Shirman, Phys. Rev. D {\bf 76},
(2007) 075004; T. Kikuchi and N. Okada, Phys. Lett. B {\bf 661},
(2008) 360; N. G. Deshpande, X. G. He, and J. Jiang, Phys. Lett. B
{\bf 656}, (2007) 91; A. Delgado, J. R. Espinosa, and M. Quiros,
JHEP {\bf 0710}, (2007) 094; K. Cheung, C. S. Li, and T. C. Yuan,
Phys. Rev. D {\bf 77}, (2008) 097701; V. Barger, Y. Gao, W. Y.
Keung, D. Marfatia, and V. N. Senoguz, Phys. Lett. B {\bf 661},
(2008) 276.

%%%%%%%%%%%%%%%%%%

%%%%%%%%%%%%%%%%%%Cosmology

\bibitem{cos} J. McDonald, arXiv: 0709.2350;
S. Hannestad, G. Raffelt, Y. Y. Y. Wong, Phys. Rev. D {\bf 76},
(2007) 121701; P. K. Das, Phys. Rev. D {\bf 76}, (2007) 123012.

%%%%%%%%%%%%%%%%%%

%%%%%%%%%%%%%%%%%%susy

\bibitem{susy} H. Zhang, C. S. Li, and Z. Li, Phys. Rev. D {\bf 76}, (2007) 116003;
Yu Nakayama, Phys. Rev. D {\bf 76}, (2007) 105009.

%%%%%%%%%%%%%%%%%%%%%%%%%
%%%%%%%%%%%%%%%%%%funda

\bibitem{funda} M. A. Stephanov, Phys. Rev. D {\bf 76}, (2007) 035008;
 M. Bander, J. L. Feng, A. Rajaraman, and Y. Shirman, Phys. Rev. D {\bf 76}, (2007) 115002;
 B. Grinstein, K. Intriligator, and I. Z. Rothstein, Phys. Lett. B {\bf 662}, (2008) 367.

%%%%%%%%%%%%%%%%%%%%%%%

%\bibitem{propagator} C. F. Chang, K. Cheung, T. C. Yuan, Phys. Lett. B {\bf 664}, (2008) 291.

%%%%%%%%%%%%%%%%%%%cross section data
\bibitem{cross-data} R. M. Davidson, Nimai C. Mukhopadhyay, and R. S.
Wittman, Phys. Rev. D {\bf 43}, (1991) 71.

%%%%%%%%%%%%%%%%%%%%%%%%%%%%other composition
\bibitem{nuclei} F. W. Stecker, Phys. Rev. {\bf 180}, (1969) 1264;
J. L. Puget, F. W. Stecker, and J. H. Bredekamp, Astrophys. J. {\bf
205} (1976) 638; W. Tkaczyk, J. Wdowczyk, and A. W. Wolfendale, J.
Phys. A {\bf 8} (1975) 1518; S. Karakula and W. Tkaczyk, Astropart.
Phys. {\bf 1} (1993) 229.

%%%%%%%%%%%%%%%%%%%%%%%%%%%%%%%our work
\bibitem{our} S. X. Chen and R. Hu, JCAP {\bf 0910}(2009) 008.

%%%%%%%%%%%%%%%%%%%%%
\bibitem{PAO-photon} J. Abraham {\it et al}., Astropart. Phys. {\bf 31}, (2009) 399;
{\it ibid}, Astropart. Phys. {\bf 29}, (2008) 243.

%%%%%%%%%%%%%%%%%%%%%%%%%%%coupling
%\bibitem{coupling} M. G. Olsson and E. T. Osypowski, Nucl. Phys. B {\bf 87}, (1975) 399;
 %Phys. Rev. D {\bf 17}, (1978) 174.

\end{thebibliography}
\end{document}